\documentclass[manuscript]{aastex}
\usepackage{natbib}
\usepackage{textcomp}
\usepackage[colorlinks=true,citecolor=blue]{hyperref}
\usepackage{graphicx,color}
\usepackage{times}
\usepackage{amssymb}

\newcommand{\nustar}{{\it NuSTAR}}
\newcommand{\swift}{{\it Swift}}
\newcommand{\phflux}{\mbox{${\rm \, ph \,\, cm^{-2} \, s^{-1}}$}}

\slugcomment{ApJ accepted}

\shorttitle{X-ray Outburst of Mrk 421}
\shortauthors{Paliya et al.}

\begin{document}

\title{Violent Hard X-ray Variability of Mrk 421 Observed by NuSTAR in 2013 April}

\author{Vaidehi S. Paliya$^{1,\,2}$, Markus B{\"o}ttcher$^{3,\,4}$, Chris Diltz$^{4}$, C. S. Stalin$^{1}$, S. Sahayanathan$^{5}$, and C. D. Ravikumar$^{2}$} 
\affil{$^1$Indian Institute of Astrophysics, Block II, Koramangala, Bangalore-560034, India}
\affil{$^2$Department of Physics, University of Calicut, Malappuram-673635, India}
\affil{$^3$Centre for Space Research, North-West University, Potchefstroom, 2520, South Africa}
\affil{$^4$Astrophysical Institute, Department of Physics and Astronomy, Ohio University, Athens, OH 45701, USA}
\affil{$^5$Astrophysical Sciences Division, Bhabha Atomic Research Centre, Mumbai-400085, India}
\email{vaidehi@iiap.res.in}

\begin{abstract}
The well studied blazar Markarian 421 (Mrk 421, $z$=0.031) was the subject of an intensive multi-wavelength campaign when it flared in 2013 April. The recorded X-ray and very high energy (VHE, E$>$100 GeV) $\gamma$-ray fluxes are the highest ever measured from this object. At the peak of the activity, it was monitored by the hard X-ray focusing telescope {\it Nuclear Spectroscopic Telescope Array} (\nustar) and \swift~X-Ray Telescope (XRT). In this work, we present a detailed variability analysis of \nustar~and \swift-XRT observations of Mrk 421 during this flaring episode. We obtained the shortest flux doubling time of 14.01$\pm$5.03 minutes, which is the shortest hard X-ray (3$-$79 keV) variability ever recorded from Mrk 421 and is on the order of the light crossing time of the black hole's event horizon. A pattern of extremely fast variability events superposed on slowly varying flares is found in most of the \nustar~observations. We suggest that these peculiar variability patterns may be explained by magnetic energy dissipation and reconnection in a fast moving compact emission region within the jet. Based on the fast variability, we derive a lower limit on the magnetic field strength of $B \ge 0.73 \delta_1^{-2/3} \, \nu_{19}^{1/3}$~G, where $\delta_1$ is the Doppler factor in units of 10, and $\nu_{19}$ is the characteristic X-ray synchrotron frequency in units of $10^{19}$~Hz. \end{abstract}
\keywords{galaxies: active --- X-rays: galaxies --- quasars: individual (Mrk\,421) --- galaxies: jets}

\section{Introduction}\label{sec:intro}
Blazars are a special class of radio-loud active galactic nuclei (AGN) whose observed broadband spectra are dominated by highly variable, non thermal, and Doppler boosted radiation from powerful relativistic jets. An exhaustive and detailed search for blazar variability at different wavelengths is necessary to understand not only the size and/or location of the emission region, but also the involved particle acceleration mechanisms and radiative processes. In this aspect, observations of extremely fast variability at very high energies (VHE, E$>$100 GeV) from some BL Lac objects \citep[e.g.,][]{2007ApJ...664L..71A} have raised questions on the radiative models proposed to explain blazar emissions. However, the lack of sensitive hard X-ray instruments prevented us to understand the connection between the observed variability at VHE, corresponding to the tail of the synchrotron self Compton (SSC) spectrum, with that at hard X-rays (synchrotron tail), under the leptonic emission scenario. Thanks to the hard X-ray mission {\it Nuclear Spectroscopic Telescope Array} \citep[\nustar,][]{2013ApJ...770..103H}, such valuable information at hard X-ray energies are now available and using them one can get deeper insights about the physics of blazar radiation zones.

Markarian 421 (Mrk 421, $z$ = 0.031) is a BL Lac object which has been studied extensively over a broad spectral range starting from radio to VHE $\gamma$-rays \citep[see e.g.,][]{2008ApJ...677..906F,2011ApJ...736..131A,2012A&A...541A.140S,2015A&A...576A.126A,2015A&A...578A..22A}. The lack of emission lines and a thermal component in its broadband spectral energy distribution (SED) suggests that the emission from the jet to be dominant rather than other external sources like the accretion disk, and the broad line region. Consistently, the high energy emission is often explained successfully by an SSC process without invoking any additional radiative component \citep[e.g.,][]{2015A&A...578A..22A}. In addition, the extension of the synchrotron spectrum up to X-rays reflects an efficient acceleration mechanism that produces a particle spectrum extending up to extremely relativistic energies. Accordingly, the SSC spectrum also extends to VHE $\gamma$-rays and hence, Mrk 421 is known to be a strong TeV emitter \citep[][]{1992Natur.358..477P}. It exhibits a flat radio spectrum, optical  polarization, and large amplitude variability throughout the electromagnetic spectrum \citep[e.g.,][]{2015A&A...576A.126A,2015MNRAS.448.3121H}. In particular, extremely fast VHE outbursts were detected where the doubling time of the flare events were found to be $<$15 minutes \citep{1996Natur.383..319G}. A positive correlation between X-rays and VHE radiation is also reported \citep[][]{2015A&A...576A.126A}, thus suggesting these emissions arise from the same region. Using the Whipple observatory ($E>400$ GeV) and All-Sky Monitor (ASM, 2$-$10 keV) onboard {\it Rossi X-ray Timing Explorer} ({\it RXTE}) data, a long term study of Mrk 421 has been performed by \citet{2014APh....54....1A} and they also report a positive correlation between VHE and X-ray fluxes.

In 2013 April, Mrk 421 underwent a giant X-ray flare and was extensively monitored by both space and ground based observational facilities \citep[e.g.,][]{2013ATel.4976....1C,2013ATel.4977....1P}, including \nustar. In this paper, using publicly available \nustar~and \swift~X-Ray Telescope (XRT) data, we perform a detailed study of the X-ray emission (0.3$-$79 keV) from Mrk 421 covering the period of this exceptional outburst. A major emphasis is given on searching for the fastest variations seen in this energy regime and also on understanding the patterns of hard X-ray variability. We use a $\Lambda$CDM cosmology with the Hubble constant $H_0=71$~km~s$^{-1}$~Mpc$^{-1}$, $\Omega_m = 0.27$, and $\Omega_\Lambda = 0.73$.

\section{Observations}\label{sec:data_red}
\subsection{\nustar}\label{subsec:nustar}
\nustar~is a hard X-ray focusing satellite consisting of two co-aligned focal plane modules referred as FPMA and FPMB\footnote{https://heasarc.gsfc.nasa.gov/docs/nustar/} \citep{2013ApJ...770..103H}. Mrk 421 was observed by \nustar~first in 2012 for pointing calibrations and later in 2013 as part of a coordinated multi-wavelength campaign \citep[][]{2013EPJWC..6104013B}. During the peak of the 2013 April outburst, \nustar~monitored Mrk 421 many times between April 10 to April 20 (MJD 56392$-$56402) and we define this as the period of interest. The \nustar~data are analyzed with the package NuSTARDAS (v.1.4.1). Calibration and cleaning of the event files are done with the task {\tt nupipeline} and using \nustar~CALDB updated on 2015 January 23. In the energy range of 3$-$79 keV, the source light curves and spectra are extracted using {\tt nuproducts}, from a circular region of 30$^{\prime\prime}$ radius centered at the position of Mrk 421, whereas the background region is selected as a circle of 70$^{\prime\prime}$ radius from a nearby region free from source contamination. To generate light curves, both FPMA and FPMB count rates are summed and background subtracted. To develop the light curves corresponding to one data point per observation ID, we select the bin size as the total duration of the observation run, whereas finer binned light curves are extracted using time bins as short as 5 minutes. In principle, one can adopt even shorter time binning, but choosing extremely short bins may result in larger flux uncertainties and a poorly constrained flux doubling time. On the other hand, a longer bins will wash out short timescale features. With this in mind, we find that 5 minute binning is optimal. Source spectra are binned to have at least 20 counts per bin and spectral fitting is performed in XSPEC \citep[][]{1996ASPC..101...17A}. 

\subsection{\swift-XRT}\label{subsec:swift}
\swift-XRT \citep{2005SSRv..120..165B} data, covering the period of the outburst, have been analyzed using the publicly available ``\swift-XRT data product generator" available at the University of Leicester website\footnote{http://www.swift.ac.uk/user\_objects/}. The details of the methodology adopted can be found in \citet{2007A&A...469..379E,2009MNRAS.397.1177E}. We extract the XRT light curves using the bin size equal to the total exposure of a particular observation, and also with 5 minute time binning. Moreover, light curves are also generated in three different energy bands with 5 minutes binning: 0.3$-$10 keV, 0.3$-$1.5 keV, and 1.5$-$10 keV. In both \nustar~and \swift-XRT analysis, we reject all the bins with $F_{\rm X}/\Delta F_{\rm X}<3$, where $\Delta F_{\rm X}$ is the associated error in the X-ray flux $F_{\rm X}$.

\section{Results}\label{sec:results}

\begin{figure*}

      \includegraphics[width=\columnwidth]{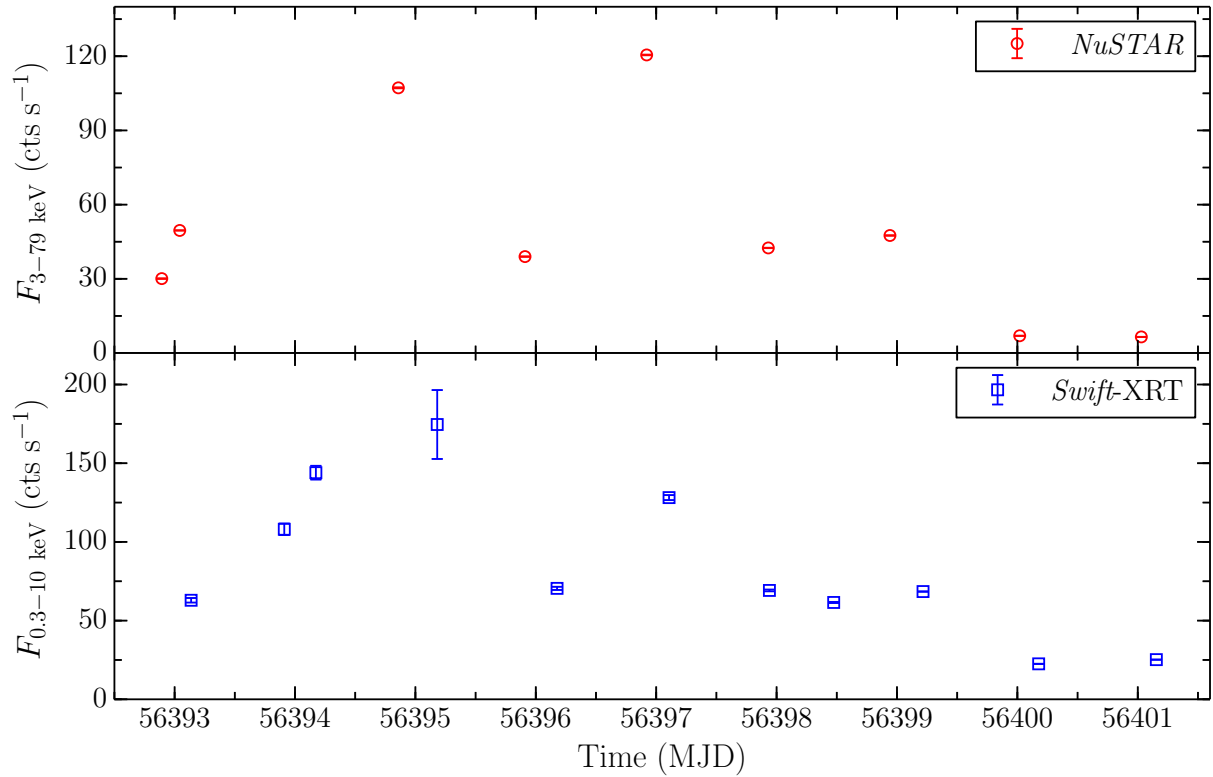}

\caption{Flux history of Mrk 421 during the period of high activity, as observed from \nustar~(3$-$79 keV, top panel) and \swift-XRT (0.3$-$10 keV, bottom panel). In both the panels, each data point corresponds to one observation ID.}\label{fig:nustar_total}
\end{figure*}

\begin{figure*}
      \includegraphics[width=\columnwidth]{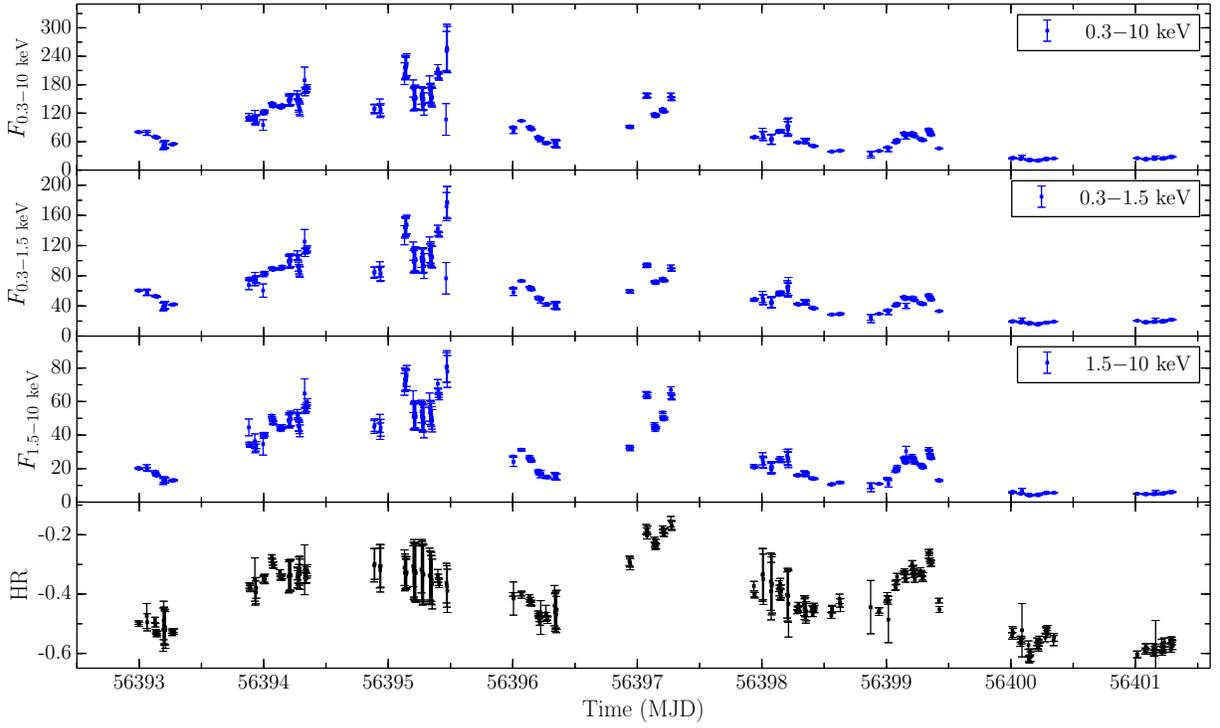}
\caption{\swift-XRT observations of Mrk 421 during the period of outburst. The selected time binning is 5 minutes and the fluxes are in units of counts s$^{-1}$. The variation of the hardness ratio (HR = $\frac{F_{\rm hard} - F_{\rm soft}}{F_{\rm hard} + F_{\rm soft}}$) is shown in the bottom panel.}\label{fig:xrt_300sec}
\end{figure*}

The \nustar~and \swift-XRT light curves of Mrk 421, covering the period of high activity, are presented in Figure~\ref{fig:nustar_total}. In this plot, each data point corresponds to one observation ID. As can be seen, two prominent flares are visible, one at around MJD 56395 and another at around MJD 56397. Further, the good photon statistics during the outburst permits us to generate light curves using shorter binning as small as 5 minutes. Generation of such shorter time binned light curves is also useful in searching for faster variability and the possible existence of patterns in the flux variations. The resultant plots are shown in the top panel of Figure~\ref{fig:nustar_300sec} and \ref{fig:xrt_300sec} for \nustar~and \swift-XRT data, respectively. Moreover, we also generate the light curves in two different energy bands, both for \nustar~(3$-$10 keV and 10$-$79 keV) and \swift-XRT (0.3$-$1.5 keV and 1.5$-$10 keV), and they are shown in the middle panels of Figure~\ref{fig:nustar_300sec} and \ref{fig:xrt_300sec}.  These observations indicate the presence of intra-day variability. Moreover, as can be seen, multiple episodes of flaring activities are observed both in soft and hard X-ray bands. For better visualization of the patterns of flux variations, 5 minute binned \nustar~light curves are also presented in Figure~\ref{fig:nustar_300sec_panel}. In this plot, each panel represents one individual \nustar~pointing. As can be seen in Figure~\ref{fig:nustar_300sec_panel}, and also in Figure~\ref{fig:nustar_300sec}, two distinct patterns are visible, a slowly varying flare and on the top of that extremely fast flux variations. Though there are also hints for similar behavior in the XRT light curves, a strong claim cannot be made due to the short exposure of the XRT observations. 

We calculate the normalized rms amplitude of variability \citep[$F_{\rm var}$,][]{2003MNRAS.345.1271V} to characterize the flux variations. It is defined as follows:
\begin{equation}
F_{\rm var} = \sqrt{\frac{S^2 - \langle \sigma_{\rm err}^2 \rangle}{\langle F
\rangle^2}}
\end{equation}
where  $\langle F \rangle$ is the mean flux, $S^2$ is the the sample variance, and $\langle \sigma_{\rm err}^2 \rangle$ is the mean-square value of the uncertainties. The error in $F_{\rm var}$ is computed as follows \citep{2008MNRAS.389.1427P,2015A&A...576A.126A},
\begin{equation}
\sigma_{F_{\rm var}}=\sqrt{F_{\rm var}^2+\sqrt{\frac{2\langle \sigma_{\rm err}^2 \rangle^2}{N\langle F\rangle^4}+\frac{4\langle \sigma_{\rm err}^2 \rangle F_{\rm var}^2}{N\langle F \rangle^2}}}-F_{\rm var}
\end{equation}
$F_{\rm var}$ is calculated for all the light curves shown in Figure~\ref{fig:nustar_300sec} and \ref{fig:xrt_300sec}. For the \nustar~observations overall in the 3$-$79 keV energy range, the $F_{\rm var}$ is found to be 0.790 $\pm$ 0.001, whereas, for the XRT data (0.3$-$10 keV) it is 0.599 $\pm$ 0.007. Due to the long exposure of each \nustar~pointing, we are able to derive $F_{\rm var}$ for individual \nustar~observation id also. The results are presented in Table~\ref{tab:nustar}. It is important to note here that the $F_{\rm var}$ for 10$-$79 keV is found to be higher than that for 3$-$10 keV in almost all the observations. Similar behavior is noticed for the \swift-XRT data where a $F_{\rm var}$ of 0.565 $\pm$ 0.006 and 0.693 $\pm$ 0.006 is obtained for the 0.3$-$1.5 keV and 1.5$-$10 keV energy ranges, respectively. Further, the fine binned light curves presented in the top panel of Figure~\ref{fig:nustar_300sec} and \ref{fig:xrt_300sec} are scanned to find the shortest flux doubling/halving time using the following formula
\begin{equation}\label{eq:flux_double}
F(t) = F(t_0).2^{(t-t_0)/\tau}
\end{equation}
where $\tau$ is the characteristic doubling/halving timescale and $F(t_0)$ and $F(t)$ are the fluxes at time $t_0$ and $t$, respectively. The conditions that the difference in fluxes at the epochs $t$ and $t_0$ is at least 3$\sigma$ significant is also met \citep[see e.g.,][]{2011A&A...530A..77F}. The shortest flux doubling time for XRT data is 64.14 $\pm$ 13.78 minutes with $\sim$5$\sigma$ significance. The absence of minute scale variability ($<15$ minutes) in fine binned XRT light curves of Mrk 421 was earlier reported by \citet{2015ApJ...802...33P}. However, the fastest flux doubling time ($t_{\rm var}$) for \nustar~observations is found to be 14.01$\pm$5.03 minutes. If a more conservative 5$\sigma$ significance is considered, then the shortest flux doubling time is 28.44$\pm$3.76 minutes. This is the shortest hard X-ray variability ever detected from Mrk 421 and is on the order of the light crossing time of the black hole's event horizon (see Section \ref{sec:conclu}). The parameters associated with this analysis are given in Table~\ref{tab:nustar}.

In the bottom panels of Figure~\ref{fig:nustar_300sec} and \ref{fig:xrt_300sec}, we show the temporal variations of the hardness ratio (HR). It is calculated using the following equation
\begin{equation}\label{eq:HR}
{\rm HR} = \frac{F_{\rm hard} - F_{\rm soft}}{F_{\rm hard} + F_{\rm soft}},
\end{equation} 
where $F_{\rm soft}$ and $F_{\rm hard}$ are soft (0.3$-$1.5 keV for XRT and 3$-$10 keV for \nustar) and hard (1.5$-$10 keV for XRT and 10$-$79 keV for \nustar) X-ray fluxes, respectively. A `harder when brighter' behavior is evident for both \nustar~and XRT light curves, especially at the peak of the flare around MJD 56397.

\begin{figure*}\label{fig:flux_index}
      \includegraphics[width=\columnwidth]{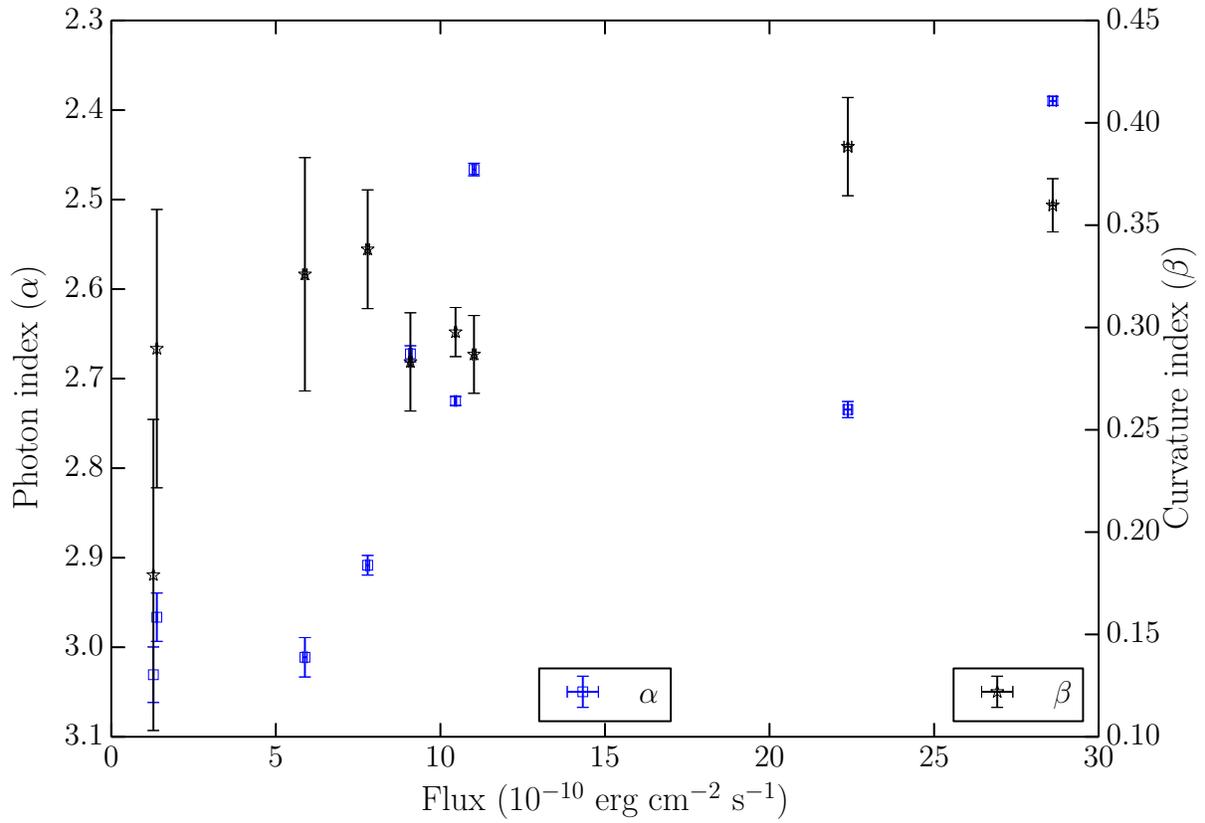}

\caption{Variations of log parabolic spectral parameters as a function of 3$-$79 keV energy flux. Left y-axis represents the photon index at the pivot energy whereas right y-axis corresponds to curvature index.}
\end{figure*}

To determine the spectral behavior of the source, we fit the \nustar~spectra with a log parabola model  \citep[see e.g.,][]{2004A&A...413..489M}, while keeping the pivot energy fixed to 10 keV. The results of the spectral fitting are presented in Table~\ref{tab:nustar_flux}. Moreover, the variations of both spectral parameters, i.e. the photon index ($\alpha$) at the pivot energy and the curvature index ($\beta$), as a function of brightness are also shown in Figure~\ref{fig:flux_index}. The spectra are found to be more curved during higher flux states, whereas there is a clear trend of `hardening when brightening', thus supporting the behavior seen in the hardness ratio plots. We note that the joint \swift-XRT and \nustar~spectral fitting, for the same period, has recently been performed by \citet{2015A&A...580A.100S} and thus it is not presented here. However, the spectral behavior of Mrk 421 observed from joint XRT-\nustar~spectral analysis, as done by \citet{2015A&A...580A.100S}, is similar to that obtained by us.
\section{Discussion and Conclusions}\label{sec:conclu}

The blazar Mrk 421 is known to exhibit fast variability at all wavelengths, especially at X-rays and VHE $\gamma$-rays \citep[][]{1996Natur.383..319G,2004ApJ...605..662C}. Though a positive correlation between these two energy bands is frequently found \citep[e.g.][]{2015A&A...576A.126A}, simultaneous hard X-ray and VHE observations were lacking during earlier measurements. The simultaneity becomes more important considering that, in leptonic models, both hard X-ray and VHE photons are expected to be produced by the same population of high energy electrons. The recent X-ray outburst of Mrk 421 was contemporaneously monitored by \nustar~and ground based Cherenkov telescopes, thus providing an excellent opportunity to constrain the radiative processes in a way that was not possible before.

Using {\it RXTE} observations of Mrk 421, \citet{2004ApJ...605..662C} reported the presence of minute scale X-ray variability, however, they could not quantify the parameters due to gaps in the data. Recently, \citet{2015ApJ...802...33P} have searched for fast X-ray variability ($<$15 minutes) among a sample of AGN monitored by the \swift-XRT, but were unsuccessful. In the energy range of 2$-$10 keV, the {\it RXTE} light curves of Mrk 421, covering the entire duration of {\it RXTE} monitoring, are publicly available\footnote{http://cass.ucsd.edu/$\sim$rxteagn/} and the details of the data reduction procedure are provided in \citet{2013ApJ...772..114R}. Using these results, we calculate the shortest flux doubling/halving time and found it to be 1.38 $\pm$ 0.37 hr. Therefore, to our knowledge, this is the first time that a statistically significant hard X-ray flux variability, as small as $\sim$14 min, has been detected from Mrk 421.

The shortest hard X-ray variability time estimated in this work is $\sim$14 min, which is similar to that observed in the VHE band by \citet{1996Natur.383..319G}. Interestingly, during the 2013 April outburst, Mrk 421 seems to show fast variability also at VHE $\gamma$-rays \citep[][]{2013ATel.4976....1C}. This observation, therefore, suggests a co-spatial origin of the X-ray and $\gamma$-ray flares. If the black hole mass of Mrk 421 is taken as 1.9 $\times$ 10$^{8}$ $M_{\odot}$ \citep[][]{{2003ApJ...583..134B}}, the observed hard X-ray variability timescale is approximately identical to the light-crossing timescale across the black hole's event horizon ($t_{\rm BH} \sim r_{\rm g}/c=GM/c^3 \sim 15$~min) which is the shortest expected variability timescale of emission powered by accretion onto the black hole. The variability timescales estimated from \nustar~observations are, thus, difficult to explain using the conventional blazar radiation models.

\begin{figure*}
      \includegraphics[width=\columnwidth]{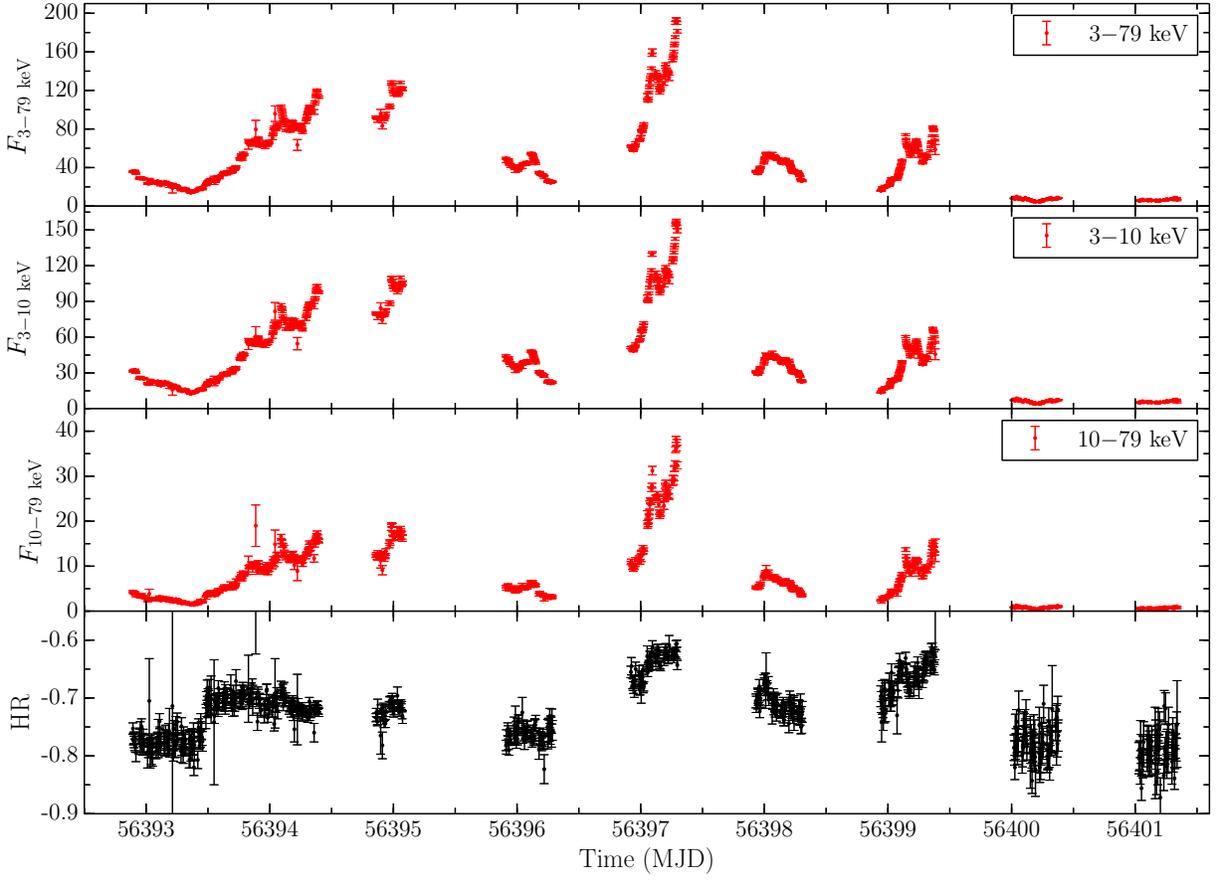}
\caption{\nustar~light curves of Mrk 421 in the energy range of 3$-$79 keV (top), 3$-$10 keV (second from top), and 10$-$79 keV (second from bottom). Bottom panel refers to the variation of hardness ratio (defined as, HR = $\frac{F_{\rm hard} - F_{\rm soft}}{F_{\rm hard} + F_{\rm soft}}$). The fluxes are in units of counts s$^{-1}$ and the adopted time binning is 5 minutes.}\label{fig:nustar_300sec}
\end{figure*}

\begin{figure*}
      \includegraphics[width=\columnwidth]{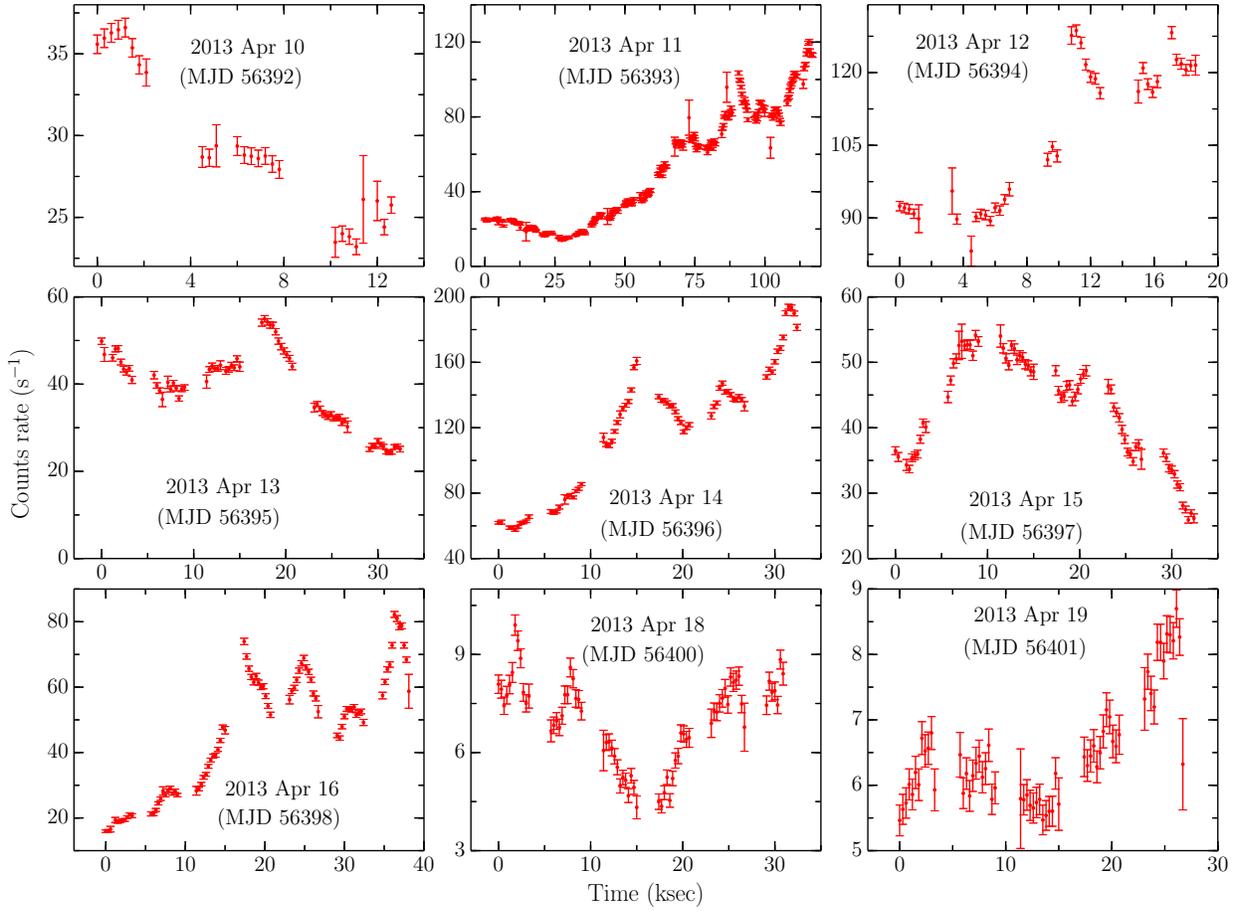}
\caption{\nustar~light curves of Mrk 421, same as in the top panel of Figure~\ref{fig:nustar_300sec}, but zoomed to show the pattern of variations . Each panel represents the \nustar~observation taken on that specific day. Other information are same as in Figure~\ref{fig:nustar_300sec}.}\label{fig:nustar_300sec_panel}
\end{figure*}

The detection of extremely fast variability from several blazars seriously challenges the commonly accepted single-zone jet models for blazar emission \citep[e.g.,][]{2007ApJ...669..862A,2011ApJ...730L...8A}. In the framework of such models, the Doppler factor of the compact emitting region has to be very high ($\gtrsim$50) in order to avoid the severe pair production of TeV photons with the synchrotron radiation and in some cases (although not in the case of the X-ray variability of Mrk 421 presented here) also to satisfy the condition $t_{\rm var} \lesssim t_{\rm BH}$ \citep[e.g.,][]{2008MNRAS.384L..19B}. However, interferometric observations of superluminal radio knots suggest lower values of the Doppler factor \citep[][]{2009AJ....138.1874L}. This apparent contradiction can be avoided by arguing that radio and hard X-ray emissions come from different emission regions and the jet is decelerated at sub-pc scales \citep[e.g.,][]{2007ApJ...671L..29L} after the production of the TeV emission. On the other hand, several alternative models have been proposed to explain such fast variability (of VHE radiation, in particular). \citet{2008MNRAS.386L..28G} have invoked the localized magneto-centrifugal acceleration of beams of electrons to explain the fast TeV variability of PKS 2155$-$304 and Mrk 501, however, this `needle' model predicts little or no variability in X-rays. The high activities seen in both VHE and X-rays from Mrk 421 during 2013 April outburst disfavors this hypothesis. \citet{2009MNRAS.395L..29G} proposed a `jet-in-a-jet' model, in which the concept of magnetic reconnection is used to explain the observed fast variability. This model not only reproduces the extremely fast TeV variations, but also predicts the observations of fast X-ray flares. The observed extremely fast hard X-ray flux variations along with the hint for high flux activity at TeV energies \citep{2013ATel.4976....1C} strengthens the hypothesis that magnetic reconnection is a possible origin of the 2013 April flare of Mrk 421. Moreover, the model of \citet{2009MNRAS.395L..29G} also predicts the presence of a slowly varying flare due to tearing of a large reconnection region. This leads to the ejection of several individual relativistic plasmoids which are thought to be responsible for fast variations. As can be seen in Figure \ref{fig:nustar_300sec} (and also in Figure \ref{fig:nustar_300sec_panel}), we do see slowly varying patterns underlying more rapid, short-term flares. This provides further support for the hypothesis of magnetic dissipation. It is interesting to note here that the above mentioned models \citep[see also][]{2012MNRAS.420..604N} have a common assumption of a small emission region moving much faster than the surrounding jet medium. A rapid flare can be observed, thus, by shorter light crossing timescale along with stronger beaming effects. On a completely different note, \citet[][and references therein]{2014MNRAS.443.3001Z} has invoked the time dependent particle injection with non-linear SSC cooling to explain the fast variability seen in the blazar light curves. 

Though the fundamental causes of the origin of the 2013 April outburst of Mrk 421 are uncertain, a few model-independent parameter estimates can be derived merely based on the assumption of a synchrotron origin of the hard X-ray emission from Mrk~421. If the radiation output of the dominant electron population is primarily by synchrotron emission, electrons of energy $\gamma \, m_e c^2$ lose energy on an observed timescale of $t_c = ([1 + z] / \delta) \, (6 \pi \, m_e c^2) / (c \, \sigma_T \, B^2 \, \gamma)$, where $\delta = 10 \, \delta_1$ is the bulk Doppler factor, $B = 1 \, B_{\rm G}$~Gauss is the magnetic field, and $\sigma_{\rm T}$ is the Thomson cross section. The electron Lorentz factor can be associated with a characteristic X-ray frequency in the \nustar~energy range, $\nu_{\rm sy} = 10^{19} \, \nu_{19} \, {\rm Hz} = 4.2 \times 10^6 \, (\delta / [1 + z]) \, B_{\rm G} \, \gamma^2$~Hz. Combining these two identities and requiring that the synchrotron cooling timescale of electrons radiating in the \nustar~regime has to be shorter than or equal to the observed minimum variability timescale, we find 
\begin{equation}
B \ge 0.73 \, \delta_1^{-2/3} \, \nu_{19}^{1/3} \; {\rm G}
\end{equation}
Thus, even for $\delta \sim 30$, the inferred magnetic field of $B \ge 0.35 \, \nu_{19}^{1/3}$~G is higher than the values of $B \lesssim 0.1$~G typically inferred from SED modeling of high frequency peaked BL Lacs such as Mrk 421. Assuming values of $B = 0.4$~G and $\delta = 30$, consistent with the above estimates, electrons radiating near the high-energy end of the \nustar~range, have Lorentz factors of 
\begin{equation}
\gamma_X \sim 4 \times 10^5 \, \left( {\delta \over 30} \right)^{-1/2} \, \left( {B \over 0.4 \, {\rm G}} \right)^{-1/2}
\, \nu_{19}^{1/2}
\end{equation}
and can produce $\gamma$-rays by Compton scattering in the Thomson regime up to photon energies of $E_{\rm T, max} \sim (\delta / [1 + z]) \, \gamma \, m_e c^2$ or 
\begin{equation}
E_{\rm T, max} \sim 6 \, \left( {\delta \over 30} \right)^{1/2} \, \left( {B \over 0.4 \, {\rm G}} \right)^{-1/2} \, 
\nu_{19}^{1/2} \; {\rm TeV}
\end{equation}
by scattering target photons of energy $E_t \sim 38 \, (\delta / 30)^{3/2} \, (B / 0.4 {\rm G})^{1/2} \, \nu_{19}^{-1/2}$~eV, i.e., UV -- soft X-ray photons. Hence, the same population of ultrarelativistic electrons can plausibly be responsible for both hard X-ray synchrotron and Compton VHE $\gamma$-ray emission, varying on comparable timescales, thus providing strong support for a leptonic (plausibly synchrotron self Compton) co-spatial origin of the X-ray and VHE $\gamma$-ray emission. 

The extremely fast variability seen at hard X-rays suggests the impulsive injection (acceleration) of electrons of the highest energies as the likely cause of the flux variations, since the highest energy electrons have the shortest cooling timescales. The injection of highly energetic particles is expected to cause not only a flux enhancement but also a spectral hardening, which is seen. However, what causes the injection of the highest energy electrons and/or what can energize the particles remains unclear. 

\acknowledgments
We are thankful to the referee for a constructive review of the manuscript. This research has made use of data, software, and/or web tools obtained from NASA’s High Energy Astrophysics Science Archive Research Center (HEASARC), a service of the Goddard Space Flight Center and the Smithsonian Astrophysical Observatory. This research has also made use of the NuSTAR Data Analysis Software (NuSTARDAS) jointly developed by the ASI Science Data Center (ASDC, Italy) and the California Institute of Technology (Caltech, USA). This work has made use of light curves provided by the University of California, San Diego Center for Astrophysics and Space Sciences, X-ray Group (R.E. Rothschild, A.G. Markowitz, E.S. Rivers, and B.A. McKim), obtained at http://cass.ucsd.edu/$\sim$rxteagn/. This work made use of data supplied by the UK Swift Science Data Centre at the University of Leicester. The work of M.B. is supported by the South African Research Chair Initiative (SARChI) of the Department of Science and Technology and the National Research Foundation\footnote{Any opinion, finding and conclusion or recommendation expressed in this material is that of the authors and the NRF does not accept any liability in this regard.} of South Africa.

\bibliographystyle{apj}
\bibliography{Master}

\begin{table*}
\footnotesize
\caption{Variability Characteristics of Mrk 421 for 5 minutes binned \nustar~Observations shown in Figure~\ref{fig:nustar_300sec}. Col.[1] and [2]: \nustar~observation date and 
observation ID; Col.[3], [4], [5]: normalized rms variability amplitude for 3$-$79 keV, 3$-$10 keV, and 10$-$79 keV 
respectively; Col.[6] and [7]: the shortest flux doubling/halving time in minutes and its significance obtained for 3$-$79 keV data; Col.[8]: 
R: rising flare D: decaying flare.}
\begin{tabular}{cccccccc}
\hline\hline
Date & OBS ID & $F_{\rm var}^{3-79~{\rm keV}}$ & $F_{\rm var}^{3-10~{\rm keV}}$ & $F_{\rm var}^{10-79~{\rm keV}}$ & $|\tau|$ & Signif. & R/D\\
~[1] & [2] & [3] & [4] & [5] & [6] & [7] & [8]\\
\hline
2013-04-10 & 60002023024 & 0.150 $\pm$ 0.005 & 0.149 $\pm$ 0.007 & 0.146 $\pm$ 0.024 & 159.56 $\pm$ 40.13 & 4.17 & D \\
2013-04-11 & 60002023025 & 0.604 $\pm$ 0.002 & 0.595 $\pm$ 0.002 & 0.642 $\pm$ 0.006 & 14.01 $\pm$ 5.03 & 3.15 & R \\
2013-04-12 & 60002023027 & 0.140 $\pm$ 0.002 & 0.137 $\pm$ 0.003 & 0.185 $\pm$ 0.008 & 47.97 $\pm$ 4.13 & 11.33& R \\
2013-04-13 & 60002023029 & 0.254 $\pm$ 0.002 & 0.223 $\pm$ 0.003 & 0.219 $\pm$ 0.007 & --               & --   & -- \\
2013-04-14 & 60002023031 & 0.320 $\pm$ 0.001 & 0.317 $\pm$ 0.002 & 0.366 $\pm$ 0.003 & 37.87 $\pm$ 4.96 & 7.63 & R \\
2013-04-15 & 60002023033 & 0.184 $\pm$ 0.008 & 0.186 $\pm$ 0.002 & 0.221 $\pm$ 0.006 & 36.46 $\pm$ 9.77 & 3.73 & D \\
2013-04-16 & 60002023035 & 0.400 $\pm$ 0.002 & 0.398 $\pm$ 0.002 & 0.451 $\pm$ 0.005 & 28.44 $\pm$ 3.76 & 7.57 & R \\
2013-04-18 & 60002023037 & 0.186 $\pm$ 0.005 & 0.182 $\pm$ 0.005 & 0.227 $\pm$ 0.015 & 20.26 $\pm$ 5.71 & 3.55 & R \\
2013-04-19 & 60002023039 & 0.123 $\pm$ 0.006 & 0.121 $\pm$ 0.006 & 0.157 $\pm$ 0.018 & --             & --   & -- \\
\hline
\end{tabular}
\label{tab:nustar}
\end{table*}

\begin{table*}
\centering
\footnotesize
\caption{Summary of the \nustar~Data Analysis}\label{tab:nustar_flux}
\begin{tabular}{cccccccc}
\tableline\tableline
 OBS ID\tablenotemark{1} & Obs. date\tablenotemark{2} & Exp.\tablenotemark{3}& $\alpha$\tablenotemark{4} & $\beta$\tablenotemark{5} & $F_{3-79~{\rm keV}}$\tablenotemark{6} & Normalization\tablenotemark{7}& Stat.\tablenotemark{8}\\
\tableline
 60002023024 & 56392.89 & 5.76  & 3.011 $\pm$ 0.022 &  0.326 $\pm$ 0.057 & 5.883  $\pm$ 0.066 & 1.250 $\pm$ 0.013 & 634.40/624\\
 60002023025 & 56393.04 & 57.51 & 2.725 $\pm$ 0.005 &  0.298 $\pm$ 0.012 & 10.460 $\pm$ 0.028 & 2.383 $\pm$ 0.006 & 1799.28/1412\\
 60002023027 & 56394.86 & 7.63  & 2.735 $\pm$ 0.009 &  0.388 $\pm$ 0.024 & 22.380 $\pm$ 0.117 & 5.217 $\pm$ 0.024 & 1084.49/1014\\
 60002023029 & 56395.90 & 16.51 & 2.908 $\pm$ 0.011 &  0.338 $\pm$ 0.029 & 7.790  $\pm$ 0.055 & 1.716 $\pm$ 0.009 & 1003.08/911\\
 60002023031 & 56396.90 & 15.61 & 2.390 $\pm$ 0.005 &  0.360 $\pm$ 0.013 & 28.605 $\pm$ 0.106 & 6.819 $\pm$ 0.020 & 1715.36/1424\\
 60002023033 & 56397.92 & 17.28 & 2.672 $\pm$ 0.009 &  0.283 $\pm$ 0.024 & 9.088  $\pm$ 0.047 & 2.080 $\pm$ 0.010 & 1036.17/1012\\
 60002023035 & 56398.93 & 20.28 & 2.466 $\pm$ 0.007 &  0.287 $\pm$ 0.019 & 11.021 $\pm$ 0.056 & 2.570 $\pm$ 0.011 & 1231.30/1188\\
 60002023037 & 56400.01 & 17.80 & 2.966 $\pm$ 0.027 &  0.290 $\pm$ 0.068 & 1.384  $\pm$ 0.015 & 0.296 $\pm$ 0.004 & 526.60/557\\
 60002023039 & 56401.02 & 15.96 & 3.031 $\pm$ 0.031 &  0.179 $\pm$ 0.076 & 1.279  $\pm$ 0.014 & 0.259 $\pm$ 0.004 & 515.62/523\\
 \tableline
\end{tabular}
\tablenotetext{1}{\nustar~observation id.}
\tablenotetext{2}{Date of observation, in MJD.}
\tablenotetext{3}{Net exposure, in ksec.}
\tablenotetext{4}{Photon index at pivot energy, in the log parabola model.}
\tablenotetext{5}{Curvature index, in the log parabola model.}
\tablenotetext{6}{Energy flux in 3$-$79 keV energy band and in units of 10$^{-10}$ erg cm$^{-2}$ s$^{-1}$.}
\tablenotetext{7}{Normalization in units of 10$^{-3}$ \phflux~keV$^{-1}$.}
\tablenotetext{8}{Statistical parameters: $\chi^2$/dof.}
\end{table*}

\newpage

\end{document}